\pdfoutput=1 
\documentclass{IEEEtran}
\usepackage{cite}
\usepackage{amsmath,amssymb,amsfonts}
\usepackage{algorithmic}
\usepackage{graphicx}
\usepackage{textcomp}
\usepackage{xcolor}
\usepackage{soul}

\usepackage{amsthm}

\usepackage{cuted} 
\usepackage{algorithm}
\usepackage{mathtools}
\usepackage{bbm}
\usepackage{caption}
\usepackage{subcaption}

\usepackage{nameref,cleveref}

\Crefname{thm}{Theorem}{Theorems}
\Crefname{prop}{Proposition}{Propositions}
\Crefname{lem}{Lemma}{Lemmas}
\Crefname{cor}{Corollary}{Corollaries}
\Crefname{defn}{Definition}{Definitions}
\Crefname{assump}{Assumption}{Assumptions}
\Crefname{conj}{Conjecture}{Conjectures}
\Crefname{alg}{Algorithm}{Algorithms}
\Crefname{appsec}{Appendix}{Appendices}

\Crefname{equation}{}{}

\Crefname{figure}{Fig.}{Figs.}
\creflabelformat{figure}{#2\textup{#1}#3}

\theoremstyle{remark}
\newtheorem{thm}{Theorem} 
\newtheorem{lem}{Lemma} 
\newtheorem{defn}{Definition} 
\newtheorem{rem}{Remark}

\def\BibTeX{{\rm B\kern-.05em{\sc i\kern-.025em b}\kern-.08em
T\kern-.1667em\lower.7ex\hbox{E}\kern-.125emX}}
\begin{document}
\title{Log-linear Dynamic Inversion Control with Provable Safety Guarantees in Lie Groups}
\author{Li-Yu Lin \IEEEmembership{Student Member, IEEE}, James Goppert \IEEEmembership{Member, IEEE}, and Inseok Hwang, \IEEEmembership{Member, IEEE}
\thanks{This work is supported in part by NASA University Leadership Initiative Project on Secure and Safe Assured Autonomy (S2A2) under Grant 80NSSC20M0161.}
\thanks{L. Lin, J. Goppert, and I. Hwang are with the School of Aeronautics and Astronautics, Purdue University, West Lafayette, IN 47906, USA (e-mail: lin1191@purdue.edu; jgoppert@purdue.edu; ihwang@purdue.edu).}}

\maketitle

\begin{abstract}
In this paper, we use the derivative of the exponential map to derive the exact evolution of the logarithm of the tracking error for mixed-invariant systems, a class of systems capable of describing rigid body tracking problems in Lie groups. Additionally, we design a log-linear dynamic inversion-based control law to remove the nonlinearities due to spatial curvature and enhance the robustness of the controller. We apply Linear Matrix Inequalities (LMIs) to bound the tracking error given a bounded disturbance amplified by the distortion matrix and leverage the tracking error bound to create flow pipes. To demonstrate the usefulness of our method, we show its application with Urban Air Mobility (UAM) scenarios using a simplified kinematic aircraft model and polynomial-based path planning methods. 

\end{abstract}


\section{Introduction}
\label{sec:introduction}
\IEEEPARstart{E}{nsuring}  the safety of an Unmanned Aerial System (UAS) in urban environments is a challenging task as a UAS operates in a dynamic environment subject to unknown uncertainties, such as wind, etc.~\cite{thipphavong2018urban}. One of the approaches to ensure the safety of a dynamical system is to construct an invariant set of system trajectories in the presence of unknown disturbances. An invariant set represents the set of trajectories in which the system remains if it starts within or enters the set~\cite{chutinan1999verification, chutinan2003computational, goppert2019security}. This notion of set invariance provides a powerful tool to analyze the worst-case impact of the unknown disturbance on the system.

For constructing invariant sets, the existing methods in the literature either use a grid-based~\cite{mitchell2005time, homer2017constrained, chen2021fastrack} or a Lyapunov function-based approach~\cite{khalil2002nonlinear, wang2016backstepping}. One of the recent methods in the grid-based approach is FaStrack~\cite{chen2021fastrack}, where the Hamilton-Jacobi (HJ) reachability of the error dynamics between a tracking system and a planning system is computed. The authors have successfully demonstrated that when the control authority is powerful, the tracking error is bounded, and the HJ partial differential equations (PDEs) converge. Another popular method for constructing invariant sets of nonlinear systems is using a Lyapunov function~\cite{khalil2002nonlinear}. A Lyapunov function can be found either by hand or using a Lyapunov-based control design methodology, such as backstepping control~\cite{kokotovic2001constructive, wang2016backstepping, tayefi2019logarithmic}. While the HJ reachability approach can compute the invariant set for a nonlinear system, it is computationally expensive. Similarly, Lyapunov-based approaches for non-linear systems can be tedious and difficult to generalize. In this work, we utilize Linear Matrix Inequalities (LMIs)~\cite{boyd1994linear} to efficiently compute the invariant set of the linearized error dynamics of the nonlinear tracking system.

For a class of nonlinear systems, group affine systems~\cite{barrau2016invariant}, whose states can be represented using Lie Groups, a unique property, the log-linear property, can be employed to convert the nonlinear structure of the Lie Group into a linear form in the Lie algebra. Modeling the states of the nonlinear system as a member of a Lie group has been successfully applied in various applications such as state estimation~\cite{barrau2016invariant}, motion control~\cite{leonard1995motion},  optimal control~\cite{bloch2008optimal, kobilarov2011discrete, colombo2020symmetry}, etc. 
In the Lie algebra, a group affine system~\cite{barrau2016invariant} behaves linearly despite nonlinearities in the original system on the Lie group, such as those arising from rigid body rotation. The Lie algebra provides a convenient coordinate system using the log-linear property for the log-linearization of error dynamics, where the number of coordinates is equal to the degrees of freedom of the dynamical system. However, the log-linear property of the error dynamics governing the deviation of a system from its reference trajectory does not hold in the presence of feedback control, disturbance, and noise, which can cause nonlinearities in the Lie algebra dynamical system. These nonlinearities are often neglected in existing work on log-linearized systems~\cite{barrau2016invariant, teng2022lie}. As a result, the exact solution is not provided, and thus the computed invariant set of the nonlinear dynamics is over-approximated.

This paper presents a generalized method for deriving the exact evolution of the log-linearization of the error dynamics for mixed-invariant systems, as defined in ~\cite{khosravian2016state}, in Lie groups using the derivative of the exponential map. While a similar method is mentioned in~\cite{li2022closed}, where they derive the log-linearization without approximation for a specific Lie group, $SE_2(3)$, our work was developed independently, we provide the exact log-linearization approach applicable to any Lie group, making our method more generic. Additionally, to enhance the control robustness, we propose a log-linear dynamic inversion-based control law. To illustrate the effectiveness of our method, we demonstrate the application in a UAM scenario example with a kinematics aircraft model in the $SE(2)$ Lie group. This example involves calculations of invariant sets and the creation of flow pipes.

In this paper, our contributions are as follows:
\begin{enumerate}
    \item We derive the exact evolution for the log-linearization of the error dynamics of mixed-invariant systems with disturbance and control input in the vehicle system for any Lie group.
    \item We design a log-linear dynamic inversion-based control law to enhance the robustness of the control system.
    \item We derive a series-form of the nonlinear input distortion matrix, $U$, for any Lie group, and compute a closed-form of $U$ for the $SE(2)$ Lie group, which is the Special Euclidean group of two-dimensional rigid body rotation and translation.
    \item We formulate flow-pipe computation for a kinematic aircraft model by using our log-linear dynamic inversion-based control strategy and applying LMIs to calculate the invariant sets.
\end{enumerate}

The rest of this paper is organized as follows. The preliminary background and problem formulation are presented in \Cref{sec:problem}. In \Cref{sec:log-linear}, we use the derivative of the exponential map in the Lie group to derive the exact log-linearization for the error dynamics of mixed-invariant systems, and design the log-linear dynamic inversion controller using the log-linearized system in the Lie algebra. In \Cref{sec:system}, we embed the two-dimensional kinematic aircraft model in the $SE(2)$ Lie group, and show an example of log-linearization in the $SE(2)$ Lie group with disturbances. In \Cref{sec:sim}, we illustrate the simulation results for our kinematic aircraft model. Finally, in \Cref{sec:conclusion}, we present our concluding remarks and immediate future works.

\section{Preliminaries and Problem Statement}
\label{sec:problem}
\subsection{Background and Preliminaries}
\begin{defn}(Mixed-Invariant System~\cite{khosravian2016state}) 
The left-invariant system~\cite{lageman2009gradient, khosravian2016state, colombo2020symmetry} in the Lie group has the form of $\dot{X} = X[l]^\wedge$, and the right-invariant system in the Lie group is $\dot{X} = [r]^\wedge X$, where $X$ is a Lie group $G$ whose associated Lie algebra is denoted $\mathfrak{g}$ and represents states of the systems. $l, r \in \mathbb{R}^n$ are inputs of the systems and may be functions of time, and $[l]^\wedge, [r]^\wedge \in \mathfrak{g}$ are elements in the Lie algebra, which define the left- and right-invariant vector fields, respectively. $[\cdot]^\wedge$ denotes the wedge operator that maps elements from $\mathbb{R}^n$ to the Lie algebra $\mathfrak{g}$. If a system is the summation of a left-invariant system and a right-invariant system, $\dot{X} = X{[l]}^{\wedge} + {[r]}^{\wedge} X $, then the system is called a \emph{mixed-invariant system}.
\end{defn}
\begin{defn} (Group Affine System \cite{barrau2016invariant, barrau2014invariant})
If a system with dynamics $\dot{X} = f(X)$ satisfies $f(A B) = f(A) B + A f(B) - A f(I) B$, then we call the system a \emph{group affine system}. Here $A$, $B$, $X \in G$, and $I$ is the identity element of the Lie Group.
\end{defn}
\begin{lem}(Mixed-Invariant systems are Group Affine~\cite{barrau2014invariant, barrau2016invariant})
Let $f(X) = X[l]^\wedge+[r]^\wedge X$, then
\begin{align*}
&f(A) B + A f(B) - A f(I) B  \\
&= (A[l]^\wedge + [r]^\wedge A)B + A(B[l]^\wedge + [r]^\wedge B)
- A([l]^\wedge + [r]^\wedge)B \\ 
&= [r]^\wedge AB + AB[l]^\wedge = f(AB) \hspace{2cm} \blacksquare
\end{align*}
\end{lem}
\begin{lem}(The Derivative of the Exponential Map~\cite{rossmann2006lie})
\label{def:dexp}
The exponential map of a matrix Lie group is a map from the matrix Lie algebra to the corresponding matrix Lie group, $\exp: \mathfrak{g} \rightarrow G$. The \emph{derivative of the exponential map} is given by:
\begin{equation}
\dfrac{d}{dt} \exp{({[\zeta(t)]}^{\wedge})} = \exp{({[\zeta(t)]}^{\wedge})} \dfrac{I - \exp{(-ad_{[\zeta]^\wedge})}}{ad_{[\zeta]^\wedge}} \dfrac{d}{dt} [\zeta(t)]^{\wedge}
\end{equation} 
Here, $I$ denotes the identity element of the group, which is the identity matrix for matrix Lie groups~\cite{rossmann2006lie, hall2015lie}, $ad_{[\zeta]^\wedge}$ is the adjoint representation of $[\zeta]^\wedge$ in the Lie algebra, and $\dfrac{I - \exp({-ad_{[\zeta]^\wedge}})}{ad_{[\zeta]^\wedge}} = \sum\limits_{k=0}^{\infty} \dfrac{(-1)^k}{(k+1)!} (ad_{[\zeta]^\wedge})^k$.
\end{lem}


Based on the log-linear property, we can log-linearize the nonlinear system with feedback control, bounded disturbance, and noise to an exact linear bounded input system using the derivative of the exponential map. The invariant set for the linear bounded input system can be found efficiently employing LMIs.

\subsection{Problem Formulation}
Consider the dynamics of two mixed-invariant vector fields, where $X \in G$ is the state of the vehicle and $\bar{X} \in G$ is the state of the reference trajectory:
\begin{equation}
\begin{aligned}
\label{eq:biX}
\dot{X} &= X{[l]}^{\wedge} + {[r]}^{\wedge} X\\
\dot{\bar{X}} &= \bar{X}{[\bar{l}]}^{\wedge} + {[\bar{r}]}^{\wedge} \bar{X}
\end{aligned}
\end{equation}
where $l, r, \bar{l}, \bar{r} \in \mathbb{R}^n$ are inputs of the systems and may be functions of time. The relationship between the two inputs is $l = \bar{l} + u_l, r = \bar{r} + u_r$. $u_l$ is the deviation of the system from the reference left-invariant vector field, $l$, and $u_r$ is the deviation of the system from the reference right-invariant vector field, $r$. The left- and right-invariant errors, $\eta_l, \eta_r \in G$ are defined as follows~\cite{barrau2016invariant}:
\begin{equation}
\begin{aligned}
\eta_l = X^{-1}\bar{X} \hspace{0.5cm} \eta_r = \bar{X}X^{-1}
\end{aligned}
\end{equation}

In this paper, we are interested in deriving the exact log-linearization for the error dynamics, $\dot{\eta_l}, \dot{\eta_r}$, of mixed-invariant systems, and designing a robust controller with provable safety guarantees.

\section{Main Result}
\label{sec:log-linear}
In this section, we first derive the exact log-linearization for the error dynamics of mixed-invariant systems for both left- and right-invariant error dynamics. We then design a log-linear dynamic inversion control law to address the nonlinear input distortion and enhance the robustness of the system.

\subsection{Exact Log-Linearization of Mixed-Invariant System}
\begin{thm}(Logarithm of Left-Invariant Error Dynamics)
For the systems in \Cref{eq:biX}, denote the left-invariant error $\eta_l$ by $\eta_l = \exp([\zeta_l]^\wedge)$, then the logarithm of the left-invariant error dynamics, $\dot{\zeta}_l$, is governed by the differential equation:
\begin{equation}
\label{eq:log_left}
\begin{aligned}
    \dot{\zeta}_l &= -ad_{[\bar{l}]^\wedge} {\zeta_l} + U_l (u_l + Ad_{X^{-1}}u_r)\\ 
    U_l &\equiv -\frac{ad_{[\zeta_l]^\wedge} \exp{(-ad_{[\zeta_l]^\wedge})}}{I - \exp{(-ad_{[\zeta_l]^\wedge})}} \hspace{0.5cm} U_l^{-1} = -\sum_{k=0}^{\infty} \frac{(ad_{[\zeta_l]^\wedge})^k}{(k+1)!}
\end{aligned}
\end{equation}
where $\zeta_l$ is the state vector of the left-invariant error dynamics in the Lie algebra, $Ad_X$ is the adjoint representation of $X$ in the Lie group, and $U_l$ is the matrix of nonlinear input distortion of inputs in the Lie algebra for the left-invariant error dynamics.
A proof is provided in \Cref{appsec:proof_left}.
\label{thm:left}
\end{thm}
\begin{thm}(Logarithm of Right-Invariant Error Dynamics)
For the systems in \Cref{eq:biX}, denote the right-invariant error $\eta_r$ by $\eta_r = \exp([\zeta_r]^\wedge)$, then the logarithm of the right-invariant error dynamics, $\dot{\zeta}_r$, is governed by the differential equation:
\begin{equation}
\label{eq:log_right}
\begin{aligned}
    \dot{\zeta}_r &= ad_{[\bar{r}]^\wedge} {\zeta_r} + U_r (u_r + Ad_{X}u_l)\\
    U_r &\equiv -\frac{ad_{[\zeta_r]^\wedge}}{I - \exp{(-ad_{[\zeta_r]^\wedge})}} \hspace{0.5cm}
    U_r^{-1} = -\sum_{k=0}^{\infty} \frac{(-1)^k (ad_{[\zeta_r]^\wedge})^k}{(k+1)!}
\end{aligned}
\end{equation}
where $\zeta_r$ is the state vector of right-invariant error dynamics in the Lie algebra, and $ad_r$ is the adjoint representation of input $r$ in the Lie algebra. $U_r$ is the matrix of nonlinear input distortion of inputs in the Lie algebra for the right-invariant error dynamics. The proof can be derived with the same process as the proof of \Cref{thm:left} in \Cref{appsec:proof_left}.
\label{thm:right}
\end{thm}

Based on \Cref{thm:left,thm:right}, we have an exact solution for log-linearization of the error dynamics of mixed-invariant systems. Since the log-linearization causes nonlinear input distortion, $U_l$ and $U_r$, the log-linear dynamics in the Lie algebra are nearly linear for a small deviation from the reference trajectory.

\begin{rem}
\label{rem:1}
It is straightforward to show from \Cref{thm:left,thm:right} that the dynamics of the left-/right-invariant error respectively are log-linear and group affine when $u_l = u_r = 0$. $\dot{\zeta}_l = -ad_{[l]^\wedge} {\zeta_l}$. $\dot{\zeta}_r = ad_{[r]^\wedge} {\zeta_r}$.
\end{rem}

\subsection{Log-Linear Dynamic Inversion Control}
For control, \Cref{thm:left,thm:right} suggest a direct approach for dynamic inversion of the log-linearized mixed-invariant vector field dynamics.
We will focus on \Cref{thm:left} but this can also be extended to \Cref{thm:right}. 
\begin{thm}(Log-Linear Dynamic Inversion Control)
Given the dynamics in \cref{eq:log_left} with $u_l \equiv u + w$ and $u_r = 0$, where $u$ represents a control input and $w$ represents a disturbance input, if $u$ is designed as:
\begin{equation}
u = U_l^{-1}B K \zeta_l 
\label{eq:control}
\end{equation}
where $B$ is the control input matrix, and $K$ is the corresponding gain matrix for the state feedback controller. 

Then, the closed-loop system becomes a linear system under the impact of magnitude bounded input disturbance, $||w||_{\infty}$. The induced matrix norm of $U_l$ can be found in~\cite{panttowards}. After substitution of \Cref{eq:control} into \Cref{eq:log_left}, the system is a linear system with bounded-input bounded-output (BIBO) stability~\cite{khalil2002nonlinear, dahleh2004lectures} given by:
\begin{equation}
\label{eq:linearsys}
\dot{\zeta}_l = (-ad_{[\bar{l}]^\wedge} + BK) \zeta_l + U_l w
\end{equation}
\label{thm:control}
\end{thm}

When using dynamic inversion control, it is important to determine if inputs are saturated in the controller. Since we use a state feedback control law, this can be efficiently checked and verified using the invariant set of the states. 

\begin{lem}(Stability of the Log-Linearized System with Log-Linear Dynamic Inversion Control Law)
Since the log-linearized system in the Lie algebra is a linear system with BIBO stability, a Lyapunov function can be found using LMIs and used to compute an invariant set to which the system is globally asymptotically stable (GAS).
\end{lem}

\begin{figure}[!t]
    \centering
    \includegraphics[width=0.8\columnwidth]{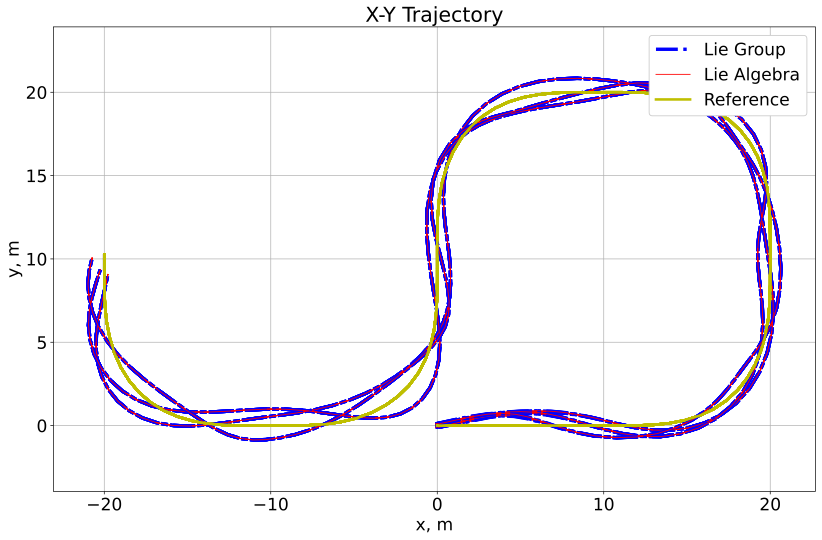}
    \caption{Lie Group-Lie Algebra Bijection for Trajectories with Control and Disturbances in $SE(2)$}
    \label{fig:LieCorrespondence}
\end{figure}

\begin{figure}[!t]
    \centering
    \begin{subfigure}[b]{0.48\textwidth}
        \includegraphics[width=0.46\columnwidth]{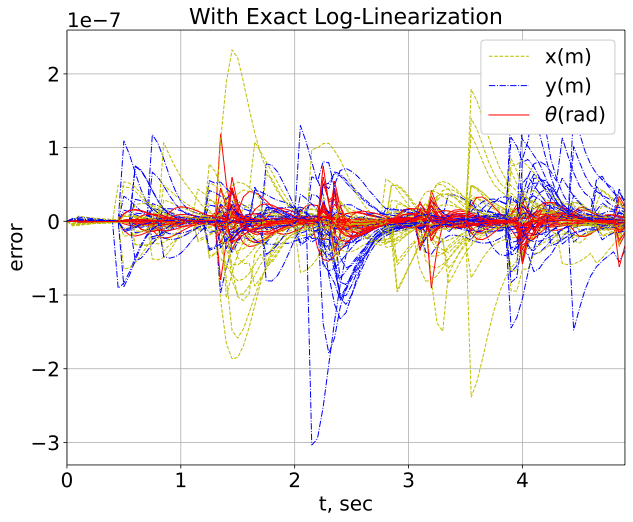}
        \includegraphics[width=0.48\columnwidth]{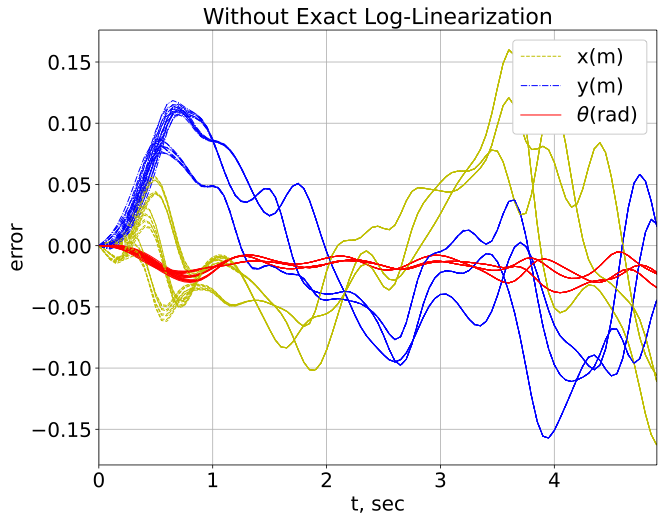}
    \end{subfigure}
    \caption{Dynamics deviation from nonlinear closed loop system in the $SE(2)$ Lie Group, Left: With Exact Log-Linearization, Right: Without Exact Log-Linearization}
    \label{fig:LieError}
\end{figure}

Figure~\ref{fig:LieCorrespondence} shows a 2D example of the bijection between the vehicle trajectories in the $SE(2)$ Lie group and the $\mathfrak{se}(2)$ Lie algebra, which are simulated with sinusoidal disturbances of fixed magnitude at varying frequencies. In order for a bijection to exist, the exponential map must be injective and surjective, which holds if the domain of the Lie algebra is restricted to $[-\pi, \pi]$~\cite{hall2015lie}. Effectively meaning that our control law will disregard multiple rotations, which is the intended behaviour. If the invariant set in the Lie algebra extends beyond the domain of $[-\pi, \pi]$, this would indicate a possibly instability through a continuous spinning motion. Thus, the LMI approach is invalidated, as the linear dynamics do not consider the angle wrapping. The figure on the left in \Cref{fig:LieError} shows that the log-linearized system dynamics in the Lie algebra do not deviate from the closed loop mixed-invariant system dynamics when the exact log-linearization is applied (the shown deviation can be arbitrarily reduced using numerical tolerances); however, there are significant deviations from the closed loop mixed-invariant system if our log-linearization is not applied (figure on the right). Therefore, we numerically demonstrate \Cref{thm:left,thm:right} by showing that a bijection exists between the trajectories in the Lie group and Lie algebra even in the presence of disturbance and control feedback, given a domain sufficiently close to the reference trajectory (e.g. avoiding angle wrapping). This is a useful result, as the Lie algebra is a vector space that is more convenient for analysis than the Lie group which is a nonlinear manifold~\cite{hall2015lie}.

\section{Example: Kinematic Aircraft Model in the $SE(2)$ Lie Group}
\label{sec:system}
In the UAM system, we consider vehicles with 3D rigid body rotation and translation, the $SE(3)$ Lie group. For constant altitude flight, this can be simplified to 2D rigid body rotation and translation, the $SE(2)$ Lie group. Since both VTOL and fixed-wing UAS dynamics can be well approximated as differentially flat~\cite{levine2009analysis, mellinger2011minimum}, we can calculate a reference trajectory that is flyable by the aircraft with negligible error in the absence of disturbance and noise. This is important as our approach focuses on the deviation of the aircraft from this reference trajectory. To simplify our approach in this paper, we consider a constant altitude kinematic model of the UAS with bounded disturbances to account for modelling error, which simplifies the UAS kinematics to a two-dimensional holonomic model. Although we are limiting our analysis to the $SE(2)$ Lie group, our algebraic approach can be generalized to other Lie groups~\cite{linapplication} (e.g., the $SE(3)$ Lie Group of rigid body rotation and translation in three dimensions).

\subsection{$SE(2)$ Lie Group Embedded Vehicle Kinematics}
Consider the kinematic aircraft model evolving on a 2D plane, as follows:
\begin{equation}
\frac{d}{dt}\theta = \omega,\; \frac{d}{dt}p_x = \cos\theta v_x - \sin\theta v_y,\; \frac{d}{dt}p_y = \sin\theta v_x + \cos\theta v_y   
\label{eq:Dubin}
\end{equation}
where $\theta \in [-\pi, \pi]$ denotes the heading angle, $p_x$ and $p_y$ denote the position. $\omega$ represents the angular velocity, $v_x$ and $v_y$ represent the forward and side translational velocity respectively. We assume our kinematic aircraft model as a holonomic model due to the side-slip angle of the aircraft and mentioned in~\cite{levine2009analysis}.

The kinematic aircraft model in~\Cref{eq:Dubin} can be embedded in the $SE(2)$ Lie group in~\Cref{eq:X}, a matrix Lie group. Special in the Special Euclidean group indicates that all matrices have determinant $1$. A Lie group is a group that is also a smooth manifold. The associated Lie algebra is a vector space tangent to the Lie group at the identity~\cite{baker2012matrix}. A Lie group and its Lie algebra are closely related, which allows calculations in one to be mapped into the other. The exponential map maps the Lie algebra to the Lie group and the logarithm map, the inverse of the exponential map, maps the Lie group to the Lie algebra~\cite{hall2015lie}.

The corresponding left-invariant system dynamics~\cite{leonard1995motion, bloch2008optimal,khosravian2016state,colombo2020symmetry} of the 2D holonomic aircraft problem, which is a mixed-invariant system with the right-invariant vector field, $r$, equal to zero, is given by:
\begin{equation}
\label{eq:X}
\dot{X} = X{[l]}^{\wedge}
\end{equation}
where $X$ represents the state of the 2D holonomic aircraft in the $SE(2)$ Lie group. Note that $l \in \mathbb{R}^3$ may be a function of time and represents the input of the vehicle, including feedback control and disturbances. The $SE(2)$ Lie group may be represented by matrices of the form:
\begin{equation}
X = \begin{bmatrix}
    R & p \\
    0 & 1                   
\end{bmatrix}
\end{equation}
where $R \in SO(2)$, which is an element of the 2D rotation group, and $p = \begin{bmatrix} p_x & p_y\end{bmatrix}^T\in \mathbb{R}^2$. The $SE(2)$ Lie group is a semi-direct product of the $SO(2)$ and $\mathbb{R}^2$~\cite{hall2015lie}. 

The corresponding Lie algebra, $\mathfrak{se}(2)$ may be represented by matrices of the form:
\begin{equation}
[x]^\wedge = \begin{bmatrix}
    \omega_\times & v \\
    0 & 0                   
\end{bmatrix}
\end{equation}
where $x = \begin{bmatrix} v & \omega \end{bmatrix}^T$ is the element in the Lie algebra. $\omega_\times$ is the corresponding skew-symmetric matrix of $\omega \in \mathbb{R}$, such that $\omega_\times = -\omega_\times^T$, $v = \begin{bmatrix} v_x & v_y \end{bmatrix}^T \in \mathbb{R}^2$. In Lie group theory, the adjoint representation~\cite{hall2015lie} transports a Lie algebra element from one tangent space to another. Here, for an element $x$ of the $\mathfrak{se}(2)$ Lie algebra, we use $ad_x$ to represent the adjoint in the Lie algebra, which can be written as:
\begin{equation}
ad_{[x]^\wedge} = \begin{bmatrix}\begin{array}{c|c}
    \omega_\times & \begin{matrix} v_y \\ -v_x \end{matrix}\\ \hline
    0 & 0
\end{array}
\end{bmatrix}
\end{equation}

\subsection{Log-Linearization in the $SE(2)$ Lie Group}
The 2D kinematics aircraft kinematics can be embedded in the $SE(2)$ Lie group as $\dot{X} = X[l]^{\wedge}$. We also consider the reference kinematics embedded in the $SE(2)$ Lie group differential equation, as 
$\dot{\bar{X}} = \bar{X}{[\bar{l}]}^{\wedge}$, where $\bar{X}$ denotes the state of the reference trajectory, and $\bar{l}$ denotes the input of the reference trajectory, and may be functions of time to account for non-trivial reference trajectories. 

As we mention in the previous section, the relationship between $l$ and $\bar{l}$ is $l = \bar{l} + u_l$, here, we consider  $u_l = u + w$, where $u$ is the feedback control input, and $w$ is the disturbance. The left-invariant error between our kinematic aircraft model and the reference kinematics is $\eta_l = X^{-1}\bar{X}$. From \Cref{thm:left} and \Cref{eq:control}, the error dynamics in the Lie algebra can be written as:
\begin{equation}
\begin{aligned}
\dot{\zeta} = (-ad_{[\bar{l}]^\wedge} + BK) \zeta + U w, \hspace{0.5cm}
U \equiv U_l
\end{aligned}
\label{eq:se2sys}
\end{equation}
where $u = U^{-1}B K \zeta$ and $\zeta \equiv \begin{bmatrix}
\zeta_x & \zeta_y & \zeta_{\theta}
\end{bmatrix}^T$.

For the $SE(2)$ Lie group, the inverse of the nonlinear input distortion matrix, $U^{-1}$, can be written in a series-form function and a closed-form matrix, which is found by reducing the sine and cosine series:
\begin{multline}
U^{-1} = -\sum_{k=0}^{\infty} \frac{(ad_{[\zeta]^\wedge})^k}{(k+1)!} \\
= 
\begin{bmatrix}
    -\frac{\sin(\zeta_{\theta})}{\zeta_\theta} & \frac{1-\cos(\zeta_\theta)}{\zeta_\theta}& -\frac{\zeta_x a+\zeta_y b}{2{\zeta_\theta}^2(\cos(\zeta_\theta)-1)}\\
    -\frac{1-\cos(\zeta_\theta)}{\zeta_\theta}& -\frac{\sin(\zeta_{\theta})}{\zeta_\theta}& \frac{\zeta_x b-\zeta_y a}{2{\zeta_\theta}^2(\cos(\zeta_\theta)-1)}\\
    0 & 0 & -1
\end{bmatrix}
\end{multline}
where $a = (-2\zeta_\theta\cos(\zeta_\theta)+2\zeta_\theta-2\sin(\zeta_\theta)+\sin(2\zeta_\theta))$ and $b = (-4\cos(\zeta_\theta)+2\cos(\zeta_\theta)+3)$. The determinant of $U^{-1}$ is $\det(U^{-1})  = \frac{2(\cos(\zeta_\theta)-1)}{\zeta_\theta^2} = -1 + \frac{\zeta_\theta^2}{12} - \frac{\zeta_\theta^4}{360} + O(\zeta_\theta^6)$, since we limit $\zeta_\theta \in [-\pi, \pi]$, the determinant will never be zero, $U^{-1}$ is always invertible. Therefore, the closed-form matrix of $U$ for the $SE(2)$ Lie group is:
\begin{equation}
\begin{aligned}
U = 
\begin{bmatrix} c & -d & \frac{\zeta_{\theta} \zeta_x \sin{\left(\zeta_{\theta} \right)} + \left(1 - \cos{\left(\zeta_{\theta} \right)}\right) \left(\zeta_{\theta} \zeta_y - 2 \zeta_x\right)}{2 \zeta_{\theta} \left(1 - \cos{\left(\zeta_{\theta} \right)}\right)}\\ 
d & c & \frac{\zeta_{\theta} \zeta_y \sin{\left(\zeta_{\theta} \right)} + \left(1 - \cos{\left(\zeta_{\theta} \right)}\right) \left(- \zeta_{\theta} \zeta_x - 2 \zeta_y\right)}{2 \zeta_{\theta} \left(1 - \cos{\left(\zeta_{\theta} \right)}\right)}\\
0 & 0 & -1\end{bmatrix}
\label{eq:U} 
\end{aligned}
\end{equation}
where $c = \frac{\zeta_{\theta} \sin{\left(\zeta_{\theta} \right)}}{2 \left(\cos{\left(\zeta_{\theta} \right)} - 1\right)}$ and $d = \frac{\zeta_{\theta}}{2}$. 

The invariant set for a system can be calculated by finding an upper bound for the singular value of the nonlinear input distortion matrix, $U$, over the invariant set. The nonlinear input distortion, $U$, depends on the error state, $\zeta=[\zeta_x,\zeta_y,\zeta_\theta]^T$. If there exists an invariant set, since the invariant set is a set of the error states, $U$ will depend on the invariant set itself. And the calculation of the invariant set also depends on the singular value of $U$, we propose a Log-linear LMI-based iterative method to compute the invariant set in \Cref{alg:iter}. An invariant set calculated by this algorithm is shown in Figure~\ref{fig:invariant3d}.

\begin{algorithm}
\caption{Calculate Log-Linear LMI-based Invariant Set}
\begin{algorithmic}[1]
\label{alg:iter}
\STATE Guess the maximum singular value, $\sigma_0$, of the nonlinear input distortion matrix, $U$, in the invariant set
\STATE Compute the invariant set using an LMI by scaling the disturbance by the maximum singular value of $U$
\STATE Calculate the maximum singular value, $\sigma_{max}$, of the U matrix over the invariant set
\STATE Assign $\sigma_{max}$ to $\sigma_0$
\STATE Repeat stpng 1-4 until converged to desired tolerance ($|\sigma_0 - \sigma_{max}| < \epsilon$), favor over-approximation for safety ($\sigma_0 > \sigma_{max}$)
\end{algorithmic}
\end{algorithm}

\section{Simulation Studies of Kinematic Aircraft Model}
\label{sec:sim}
In this section, we consider an illustrative UAM scenario where a fixed-wing aircraft flies along a reference trajectory at a constant altitude. The reference trajectory is generated by polynomial trajectory planning~\cite{mellinger2011minimum, richter2016polynomial, lynch2017modern}. We consider a kinematic model of the fixed-wing aircraft. Our model is given by the equations in \Cref{eq:Dubin}, which can be embedded in the $SE(2)$ Lie group, \Cref{eq:X}. The log-linear error dynamics between the vehicle model and the reference model in the Lie algebra are given in \Cref{eq:se2sys}. For the external disturbances in the $\theta$ direction, we assume $||w_2||_\infty = 0.1$ rad/s. For disturbances in the $\{x, y\}$ direction, we consider two different disturbance magnitudes. We assume the magnitude of the bounded disturbance inputs, $||w_1||_\infty$, as $1$ m/s and $5$ m/s, respectively.

\subsection{Invariant Set}
\begin{figure}[!t]
    \centering
    \begin{subfigure}[b]{0.45\textwidth}
        \includegraphics[width=0.9\columnwidth]{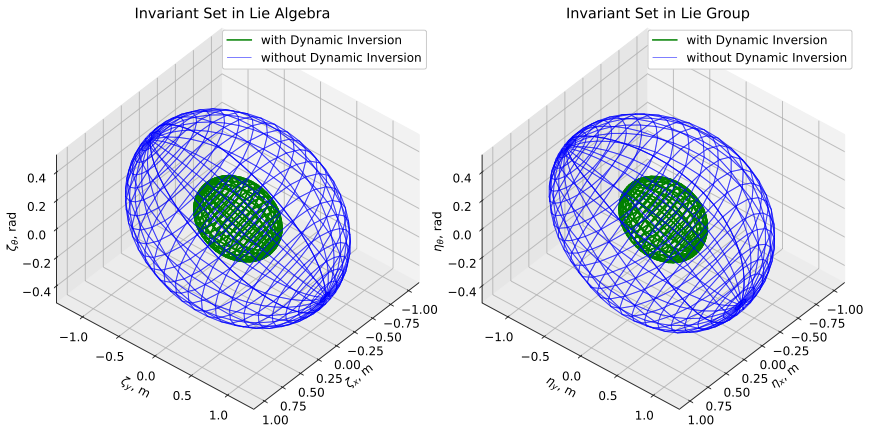}
    \end{subfigure}
    \begin{subfigure}[b]{0.45\textwidth}
        \includegraphics[width=0.9\columnwidth]{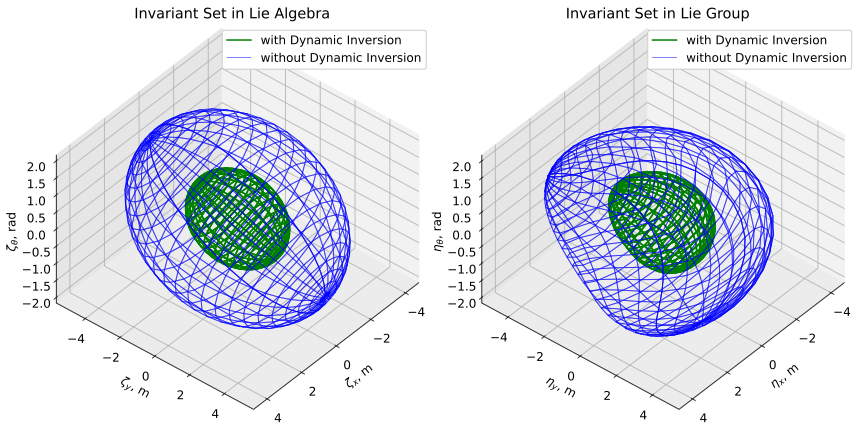}
    \end{subfigure}
    \caption{Invariant Set Comparison in three dimensions with and without Dynamic Inversion, Left: Invariant Set in Lie Algebra, Right: Invariant Set in Lie Group, Top: Small Disturbance, Bottom: Large Disturbance}
    \label{fig:invariant3d}
\end{figure}

Since the inputs of the reference trajectory may be functions of time, we use the LMI to calculate the invariant set for a linear polytopic system with bounded disturbances~\cite{boyd1994linear}. The bounds of the inputs to calculate the invariant set are based on the bounds of the reference inputs, where $\bar{v}_x \in [18, 20]$ m/s, $\bar{v}_y = 0$ m/s, and $\bar{\omega} \in [-\pi/2, \pi/2]$ rad/s. We use the Linear-Quadratic-Regular (LQR) method to find the feedback controller for the log-linearized system in the $\mathfrak{se}(2)$ Lie algebra. After employing our iterative method, \Cref{alg:iter}, to find the desired maximum singular value of $U$, the final result of the invariant set in the Lie algebra and the Lie group is shown in \Cref{fig:invariant3d}. The figures show the invariant set with small initial states in the $SE(2)$ Lie group, $[x_0, y_0, \theta_0] = [0.1, 0.1, \pi/100]$. The invariant set in the Lie group is constructed by applying the exponential map from the set in the Lie algebra. 

If the control input does not employ dynamic inversion, the control input $u$ in \Cref{eq:control} becomes $BK\zeta$, and the error dynamics in the Lie algebra in \Cref{eq:se2sys} can be rewritten as:
\begin{equation}
\dot{\zeta} = (-ad_{[\bar{l}]^\wedge} + BK)\zeta + (U-I)BK\zeta + Uw
\end{equation}
Here, we use the $SE(2)$ example to compare the difference between the two controllers. For the control input without dynamics inversion, we consider the nonlinear term of the control input as a bounded input in the LMI calculation. Blue lines in \Cref{fig:invariant3d} show the result of the invariant set for the controller without dynamic inversion. Since the nonlinear term of the control input is considered a bounded input, the invariant set has a larger tracking error bound than the controller with dynamic inversion. Comparing the invariant sets in \Cref{fig:invariant3d}, we can see that the invariant set is approximately twice as large in the case of the control input without dynamic inversion compared to the case with dynamic inversion. Therefore, the invariant set has a smaller tracking error bound with our designed controller.  

\begin{figure}[!t]
    \centering
    \begin{subfigure}[b]{0.45\textwidth}
        \includegraphics[width=0.49\columnwidth]{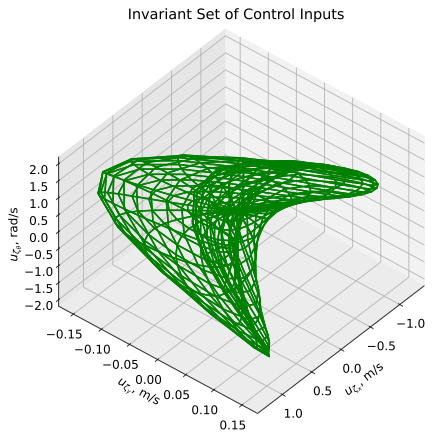}
        \includegraphics[width=0.49\columnwidth]{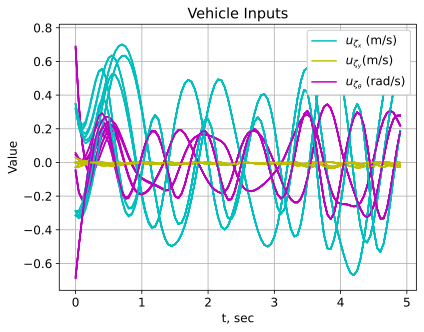}
    \end{subfigure}
    \caption{Left: Saturation Bound of Control Inputs, Right: Time History Plot of Control Inputs}
    \label{fig:control}
\end{figure}

The Lyapunov function in the LMI approach is an ellipsoid, which is the shape of the invariant set for the log-linearized system in the $\mathfrak{se}(2)$ Lie algebra as shown on the left of \Cref{fig:invariant3d} with green lines. The invariant sets in the $SE(2)$ Lie group, which are mapped from the invariant sets in the $\mathfrak{se}(2)$ Lie algebra using the exponential map, are shown on the right of \Cref{fig:invariant3d} with green lines. Since the invariant set in the $SE(2)$ Lie group is not an ellipsoid after mapping, as shown in \Cref{fig:invariant3d}, and is in the vehicle body frame; therefore, the invariant set must be rotated along the trajectory to create flow pipes. The left figure in \Cref{fig:control} shows the invariant set of our log-linear dynamic inversion control inputs for the small disturbance case, the saturation bounds for each direction are $u_{\zeta_x} \in [-1.25, 1.25]$ m/s, $u_{\zeta_y} \in [-0.16, 0.16]$  m/s, and $u_{\zeta_\theta} \in [-1.92, 1.92]$ rad/s. The right figure in \Cref{fig:control} shows the time history plot of the control inputs for the simulated trajectories with sinusoidal and square wave disturbances in the small disturbance case, and each control input stays within the saturation bound for all time.

\subsection{Flow Pipes}
To verify the safety properties for the UAM mission, we construct flow pipes along a reference trajectory. Any Lyapunov function will be sufficient for our method and we follow a general approach of the flow pipes creation in~\cite{goppert2019security}. To construct the flow pipe, we first propagate the reference trajectory using polynomial trajectory planning for a fixed time interval and compute the interval hull of the reference trajectory. The interval hull is the smallest box enclosing the set for the propagation duration. Since our invariant set depends on the direction of the reference trajectory, we sweep the invariant set from the minimum rotational angle to the maximum rotational angle in that time interval. Using the swept invariant set and interval hull, we then compute the convex hull for the flow pipe segment using the Minkowski sum~\cite{varadhan2004accurate}. Therefore, the flow pipe for the trajectory can be constructed by repeating the above processes for the next time interval. 

\begin{figure}[t]
    \centering
    \begin{subfigure}[b]{0.5\textwidth}
        \includegraphics[width=0.49\columnwidth]{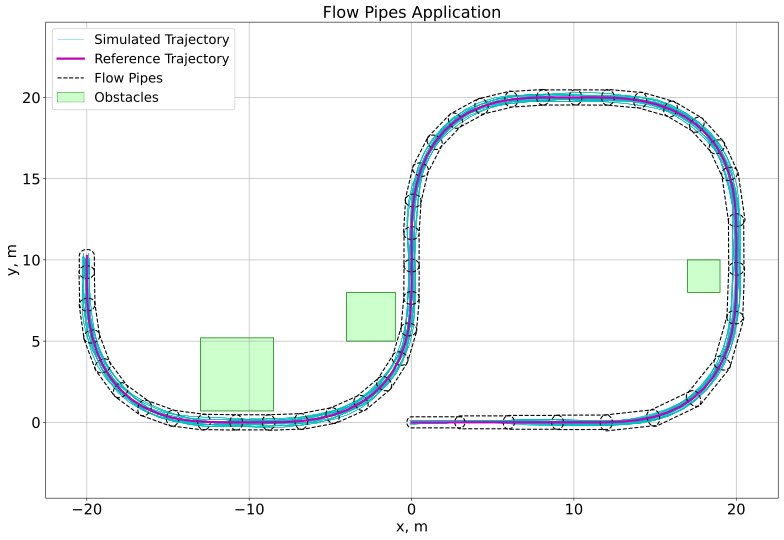}
        \includegraphics[width=0.49\columnwidth]{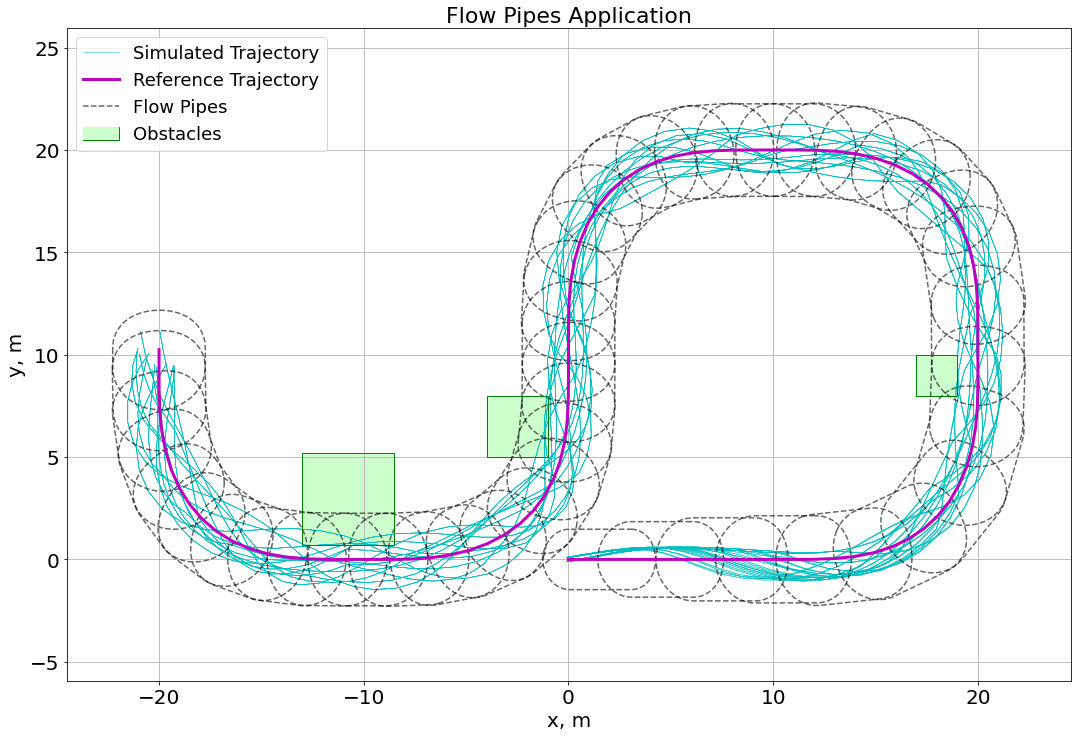}
    \end{subfigure}
    \caption{Flow Pipes Applications for the UAM Problem, Left: Small Disturbances, Right: Large Disturbances}
    \label{fig:flowpipes_application}
\end{figure}

Figure~\ref{fig:flowpipes_application} shows the application of flow pipes to the UAM problem. The green squares represent collision zones or obstacles (e.g., buildings) that the vehicle needs to avoid. Here, we simulate both sinusoidal and square wave wind disturbances. We simulate two different scenarios for the vehicle. The first scenario is simulated with small wind disturbances with a magnitude of $1$ m/s, which is shown at the top of \Cref{fig:flowpipes_application}. We can see that the green collision zones, obstacles, and the flow pipes do not intersect; therefore, we can confirm the safety of the vehicle, since the simulation shows that the vehicle won't hit the buildings during the mission. Another scenario is simulated with large wind disturbances with a magnitude of $5$ m/s, which is shown at the bottom of \Cref{fig:flowpipes_application}. We can see that the buildings and the flow pipes are now intersecting. Therefore, we are able to confirm that it is not safe for the vehicle to conduct the mission with wind disturbances of this magnitude. We can also see that some simulated trajectories collide with obstacles and the mission is therefore unsafe.

\section{Conclusion}
\label{sec:conclusion}
In this paper, we presented a novel approach to derive the exact log-linearization of the error dynamics of mixed-invariant systems. This approach employed the derivative of the exponential map in Lie group theory. Our analysis of the evolution of the log-linearized system yielded a nonlinear input distortion matrix. We were able to leverage the singular value of this bounded input distortion matrix to reformulate our tracking error dynamics to a linear system with bounded-input bounded-output (BIBO) stability. Additionally, we designed a new dynamic inversion-based control law to simplify the log-linearized tracking error dynamics to a linear system with a bounded disturbance and enhanced the robustness of the system. In the simulation, we demonstrated the application of our algorithm to safety verification in an Urban Air Mobility (UAM) scenario. We also compared the bound of the invariant set with and without the dynamic inversion-based control law.

As future work, we intend to consider the invariant set as time-varying to narrow the bound of the flow pipes and extend our approach to more complex aircraft dynamic models with the $SE_2(3)$ and $SE(3)$ Lie groups.

\appendices
\crefalias{section}{appsec}

\section{Proof of \Cref{thm:left}}
\label{appsec:proof_left}

Using the product rule, the left-invariant error dynamics, which are the derivative of $\eta_l$, can be written as: 
\begin{equation}
\begin{aligned}
\dot{\eta}_l &= -X^{-1}\dot{X}X^{-1}\bar{X} + X^{-1}\dot{\bar{X}} \\
&= \eta_l{[\bar{l}]}^{\wedge} - {[\bar{l}+u_l]}^{\wedge} \eta_l - Ad_{X^{-1}}{[u_r]^{\wedge}} \eta_l 
\label{eq:deta_l} 
\end{aligned}
\end{equation}
Let $\eta_l = \exp({[\zeta_l]}^{\wedge})$, where $\zeta_l$ is the left-invariant error in the Lie algebra. Based on \Cref{def:dexp}, the derivative of the exponential map can be written as:
\begin{equation}
[\dot{\zeta}_l]^{\wedge} = \dfrac{ad_{[\zeta_l]^\wedge}}{I - \exp{(-ad_{\zeta_l)}}} \eta_l^{-1} \dot{\eta}_l
\label{eq:exp_eta_l}
\end{equation}
The left-invariant error dynamics in \Cref{eq:deta_l} can be rewritten as:
\begin{multline}
\eta_l^{-1} \dot{\eta}_l = [\bar{l}]^{\wedge} - Ad_{{\eta_l}^{-1}} [\bar{l} + u_l]^{\wedge} - Ad_{{\eta_l}^{-1}} Ad_{X^{-1}}[u_r]^{\wedge} \\
= \left(I - \exp{(-ad_{[\zeta_l]^\wedge})}\right)[\bar{l}]^{\wedge}  - \exp{(-ad_{[\zeta_l]^\wedge})}[u_l]^{\wedge} \\ 
- \exp{(-ad_{[\zeta_l]^\wedge})}Ad_{X^{-1}}[u_r]^{\wedge} 
\label{eq:etal-1eta}  
\end{multline}

The corresponding dynamics for left-invariant error in the Lie algebra can be derived as:
\begin{equation}
\begin{aligned}
[\dot{\zeta}_l]^{\wedge} &= \frac{ad_{[\zeta_l]^\wedge}}{I - \exp{(-ad_{[\zeta_l]^\wedge})}} [\left(I - \exp{(-ad_{[\zeta_l]^\wedge})}\right)[\bar{l}]^{\wedge} \\
&- \exp{(-ad_{[\zeta_l]^\wedge})}[u_l]^{\wedge} - \exp{(-ad_{[\zeta_l]^\wedge})}Ad_{X^{-1}}[u_r]^{\wedge}]\\
&= -ad_{[\bar{l}]^\wedge} [{\zeta_l}]^{\wedge}\\  &-\frac{ad_{[\zeta_l]^\wedge} \exp{(-ad_{[\zeta_l]^\wedge})}}{I - \exp{(-ad_{[\zeta_l]^\wedge})}} ([u_l]^{\wedge} + Ad_{X^{-1}}[u_r]^{\wedge})  
\end{aligned}
\label{eq:zeta-dot-l-wedge}
\end{equation}
$\frac{ad_{[\zeta_l]^\wedge} \exp{(-ad_{[\zeta_l]^\wedge})}}{I - \exp{(-ad_{[\zeta_l]^\wedge})}}$ can be expressed as a matrix power series of $ad_{[\zeta_l]^\wedge}$, and $ad_{[\zeta_l]^\wedge}$ is a linear operator on the Lie algebra, therefore, we can use the Lie vee operator, ${[\cdot]}^{\vee}$, which is the inverse of the wedge operator, to map the error dynamics from a form of elements of the Lie algebra \Cref{eq:zeta-dot-l-wedge} to elements of the vector space $\mathbb{R}^n$ \Cref{eq:log_left} as shown in \Cref{thm:left}. 

Let $U_l \equiv -\frac{ad_{[\zeta_l]^\wedge} \exp{(-ad_{[\zeta_l]^\wedge})}}{I - \exp{(-ad_{[\zeta_l]^\wedge})}}$, since each term in this expression is a power of $ad_{[\zeta_l]^\wedge}$, multiplication is commutative, thus, $U_l^{-1} = -\frac{I - \exp{(-ad_{[\zeta_l]^\wedge})}}{ad_{[\zeta_l]^\wedge} \exp{(-ad_{[\zeta_l]^\wedge})}}$, using the taylor series of exponential, we can find a series form for $U_l^{-1}$:
\begin{equation}
\begin{aligned}
    U_l^{-1} &= -\frac{I - \exp{\left(-ad_{[\zeta_l]^\wedge}\right)}}{ad_{[\zeta_l]^\wedge} \exp{\left(-ad_{[\zeta_l]^\wedge}\right)}}\\
    &= -\frac{1}{ad_{[\zeta_l]^\wedge}} \left({\exp\left({-ad_{[\zeta_l]^\wedge}}\right)}^{-1} - I\right) \\
    &= -\frac{1}{ad_{[\zeta_l]^\wedge}} \left(\exp{\left({ad_{[\zeta_l]^\wedge}}\right)}-I\right) \\
    &= -\frac{1}{ad_{[\zeta_l]^\wedge}} \left(\sum_{m=0}^{\infty} \frac{\left(ad_{[\zeta_l]^\wedge}\right)^m}{m!} -I\right)\\
    &= -\sum_{m=1}^{\infty} \frac{\left(ad_{[\zeta_l]^\wedge}\right)^{m-1}}{m!} = -\sum_{k=0}^{\infty} \frac{\left(ad_{[\zeta_l]^\wedge}\right)^k}{\left(k+1\right)!}
\end{aligned}
\end{equation}

\bibliographystyle{IEEEtran}
\bibliography{ref}

\end{document}